\documentclass[aps,prl, twocolumn,superscriptaddress,showpacs]{revtex4}

\usepackage{color}
\usepackage{graphicx}
\usepackage{natbib}
\usepackage{amsmath}

\begin{document}

\newcommand{\mycomment}[1]{\textbf{\textcolor{red}{#1}}}
\newcommand{\bra}[1]{\langle #1|}
\newcommand{\ket}[1]{|#1\rangle}
\newcommand{\braket}[2]{\langle #1|#2\rangle}

\title{Heralded Generation of Ultrafast Single Photons in Pure Quantum States}
\author{Peter J. Mosley}
\email{p.mosley1@physics.ox.ac.uk}
\affiliation{Clarendon Laboratory, University of Oxford, Parks Road, Oxford, OX1 3PU, UK}
\author{Jeff S. Lundeen}
\affiliation{Clarendon Laboratory, University of Oxford, Parks Road, Oxford, OX1 3PU, UK}
\author{Brian J. Smith}
\affiliation{Clarendon Laboratory, University of Oxford, Parks Road, Oxford, OX1 3PU, UK}
\author{Piotr Wasylczyk}
\affiliation{Clarendon Laboratory, University of Oxford, Parks Road, Oxford, OX1 3PU, UK}
\affiliation{Institute of Experimental Physics, Warsaw University, ul. Hoza 69, 00-681 Warsaw, Poland}
\author{Alfred B. U'Ren}
\affiliation{Division de Fisica Aplicada, Centro de Investigacion Cientifica y de Educacion Superior de Ensenada, Ensenada, Baja California, 22800 Mexico}
\author{Christine Silberhorn}
\affiliation{Max--Planck--Forschungsgruppe fur Optik, Gunther-Scharowsky Str. 1/Bau 24, 91058 Erlangen, Germany}
\author{Ian A. Walmsley}
\affiliation{Clarendon Laboratory, University of Oxford, Parks Road, Oxford, OX1 3PU, UK}


\begin{abstract}
We present an experimental demonstration of heralded single photons prepared in pure quantum states from a parametric downconversion source. It is shown that, through controlling the modal structure of the photon pair emission, one can generate pairs in factorable states and thence eliminate the need for spectral filters in multiple-source interference schemes. Indistinguishable heralded photons were generated in two independent spectrally engineered sources, and, by performing a Hong-Ou-Mandel interference between them without spectral filters at a raw visibility of 94.4\%, their purity was measured to be over 95\%.
\end{abstract}

\maketitle

Optical quantum information processing (QIP) either by linear optical gates \cite{Knill2001A-scheme-for-efficient-quantum, OBrien2003Demonstration-of-an-all-optical-quantum, Pittman2005Experimental-demonstration-of-a-quantum} or by the formation of cluster states \cite{Browne2005Resource-Efficient-Linear-Optical, Lu2007Experimental-entanglement-of-six-photons} relies on the coalescence of two independent photons incident on a beamsplitter. This Hong-Ou-Mandel interference (HOMI) \cite{Hong1987Measurement-of-subpicosecond-time} effect will only occur completely if the two photons are both pure and indistinguishable \cite{Bovino2005Direct-Measurement-of-Nonlinear, Rohde2005Frequency-and-temporal-effects}. There are two common approaches to the generation of single photons: cavity quantum electrodynamics (CQED) in which a single emitter such as an atom \cite{Boozer2007Reversible-State-Transfer, Hijlkema2007A-single-photon-server-with}, ion \cite{Maunz2007Quantum-interference-of-photon}, quantum dot \cite{Santori2002Indistinguishable-photons-from}, or crystal vacancy \cite{Jacques2007Experimental-Realization-of-Wheelers} is placed in a resonant optical cavity, and photon pair generation in a bulk medium, for example by four-wave mixing in a fiber \cite{Fulconis2007Nonclassical-Interference-and-Entanglement} or by three-wave mixing in a nonlinear crystal \cite{Hong1987Measurement-of-subpicosecond-time}. However, it has not been proven that single photons generated by either technique can fulfill both requirements of purity and indistinguishability \cite{Santori2002Indistinguishable-photons-from, Hijlkema2007A-single-photon-server-with, Fulconis2007Nonclassical-Interference-and-Entanglement, Maunz2007Quantum-interference-of-photon}.

In this Letter we show that single photons derived from pair generation by parametric downconversion (PDC) in a birefringent nonlinear crystal can be prepared directly in pure states. Usually, this purity is limited by correlations within each pair arising from the constraints of energy and momentum conservation between the pump and daughter fields. Hence photon pairs are emitted into many highly correlated spatio-temporal modes and, due to these quantum correlations, the act of detecting one photon (known as heralding) projects the remaining single photon into a mixed state. This renders it useless for most quantum information processing applications. A common strategy is to discard photons from the mixed state by spectral filtering to approximate a single mode \cite{Riedmatten2003Quantum-interference-with, Kaltenbaek2006Experimental-Interference-of-Independent, Lu2007Experimental-entanglement-of-six-photons}. This has the drawback that the purity of such a sub-ensemble only approaches unity as the filter bandwidth tends to zero. Hence, to achieve high purity one must use narrowband filters, making the generation rate of photons in useable modes prohibitively low.

\begin{figure}
\includegraphics[width= 0.48 \textwidth]{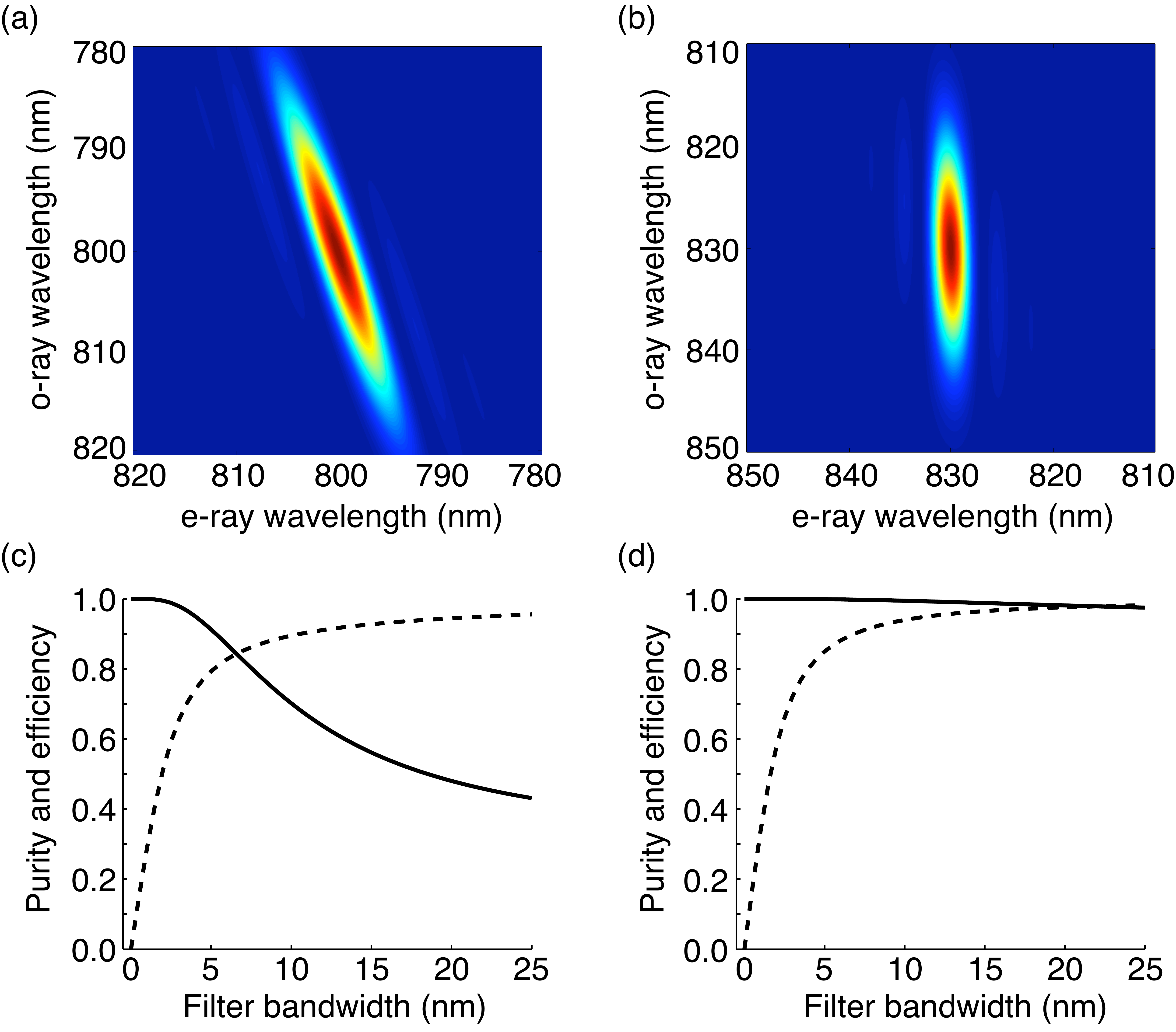}
\caption[The advantages of a factorable two-photon state.]{(Color online) A numerically simulated comparison of collinear type-II pair generation in a 2\,mm $\beta$-barium-borate (BBO) crystal pumped at 400\,nm ((a) and (c)) and a 5\,mm potassium-dihydrogen-phosphate (KDP) crystal pumped at 415\,nm ((b) and (d)). (a) and (b) show the two-photon joint spectral probability distributions. Theoretical optimum heralding efficiency (dashed) and purity of the heralded single photons (solid) are plotted in (c) and (d) as a function of filter bandwidth. We note that if we require a purity of 95\%, for BBO the maximum possible heralding efficiency is necessarily reduced to 0.75, whereas for KDP this purity does not set an upper bound on the heralding efficiency as no spectral filtering is required. In both cases, the pump was modeled as plane wave with a FWHM spectral bandwidth of 4\,nm.}
\label{fig:1}
\end{figure}

For any spectrally-filtered source of photon pairs, the tradeoff between purity and count rate imposes a fundamental limit on the production rate of heralded photons and, in the situation where \textit{both} arms are filtered, also the heralding efficiency (see Fig.\,\ref{fig:1}). However, in an ideal heralded single photon source all three parameters --- purity, count rate, and heralding efficiency --- must be high. If the purity is low then the fidelity of any operation will be poor, yet on the other hand, limited production rates result in any protocol taking a prohibitively long time \cite{Halder2007Entangling-independent-photons}. Simply reducing the filter bandwidth and increasing the pump power does not solve this problem. As one pumps harder, the background contribution from simultaneous generation of multiple pairs increases, reducing the visibility of any interference effects \cite{Fulconis2007Nonclassical-Interference-and-Entanglement, Halder2007High-coherence-photon}. Furthermore, low heralding efficiencies combined with random fluctuations in the properties of consecutive photons from filtered sources have forced all previous implementations of optical QIP protocols to remain in the post-selected regime --- successful gate operations must be conditioned on the presence of the logical qubits. This problem of post-selection is a well-known obstacle to scaling up QIP schemes \cite{Braunstein1998A-posteriori-teleportation}.

Additionally, the problem of spectral filtering to eliminate correlations becomes more extreme as one moves to the production of higher photon numbers in each mode. Loss quickly degrades entanglement in photon number, required for example in continuous variable entanglement distillation. Here, the need for high-visibility interference between photons from independent sources as well as number entanglement within the emission from each source demands that the photon pairs are generated initially in states that are uncorrelated in all degrees of freedom other than number \cite{Lvovsky2007Decomposing-a-pulsed-optical}.

It has been shown theoretically that by careful control of the modes available for PDC, each pair of daughter photons can be free from any correlations in their spatio-temporal degrees of freedom \cite{Grice2001Eliminating-frequency-and-space-time}. The PDC emission modes are restricted by adapting the idea of vacuum mode engineering, well known in CQED, to the case of bulk media and free space propagation. In CQED the cavity enclosing the single emitter ensures that it will radiate only into a single mode, however, a bulk medium contains many emitters. The coherent phasing of these determines the emission modes, which are are generally numerous and highly correlated \cite{Walton2004Generation-of-polarization-entangled-photon, Torres2005Indistinguishability-of-entangled-photons, Hodelin2006Optimal-generation-of-pulsed}. It is the multimode nature of the resulting fields that inevitably produces heralded photons in mixed states. In contrast, by choosing a nonlinear medium with the correct dispersion characteristics, we reduce the number of modes into which the downconverted fields can couple efficiently. Hence photon pairs can be created in uncorrelated states resulting in pure heralded single photons.

The entangled state emitted by a single type-II downconverter can be written as:
\begin{equation}
\ket{\Psi(\omega_{\text{e}},\omega_{\text{o}})} = \, \ket{0} \, + \eta \iint d\omega_{\text{e}} d\omega_{\text{o}} \, f(\omega_{\text{e}},\omega_{\text{o}}) a^{\dagger}_{\text{e}}(\omega_{\text{e}}) a^{\dagger}_{\text{o}}(\omega_{\text{o}}) \ket{0},
\label{eq:pdc_state}
\end{equation}
up to the two-photon component and considering explicitly only the frequencies, $\omega_{\text{e}}$ and $\omega_{\text{o}}$, of the two photons. The probability of creating a pair is $|\eta|^{2}$, and $a^{\dagger}_{\text{e}}(\omega_{\text{e}})$ and $a^{\dagger}_{\text{o}}(\omega_{\text{o}})$ are the creation operators for photons in polarization modes e and o. The purpose of heralding is to eliminate the vacuum component of Eq.\,(\ref{eq:pdc_state}), thus projecting the remaining photon into a state whose purity is determined by the factorability of the joint spectral amplitude function, $f(\omega_{\text{e}},\omega_{\text{o}})$ \cite{Grice2001Eliminating-frequency-and-space-time}.  Hence it is a necessary condition for pure heralded photons that the joint amplitude be factorable, $f(\omega_{\text{e}},\omega_{\text{o}}) = f_{e}(\omega_{\text{e}}) f_{o}(\omega_{\text{o}})$, so that only one spectral mode is present.

\begin{figure}
\includegraphics[width= 0.48 \textwidth]{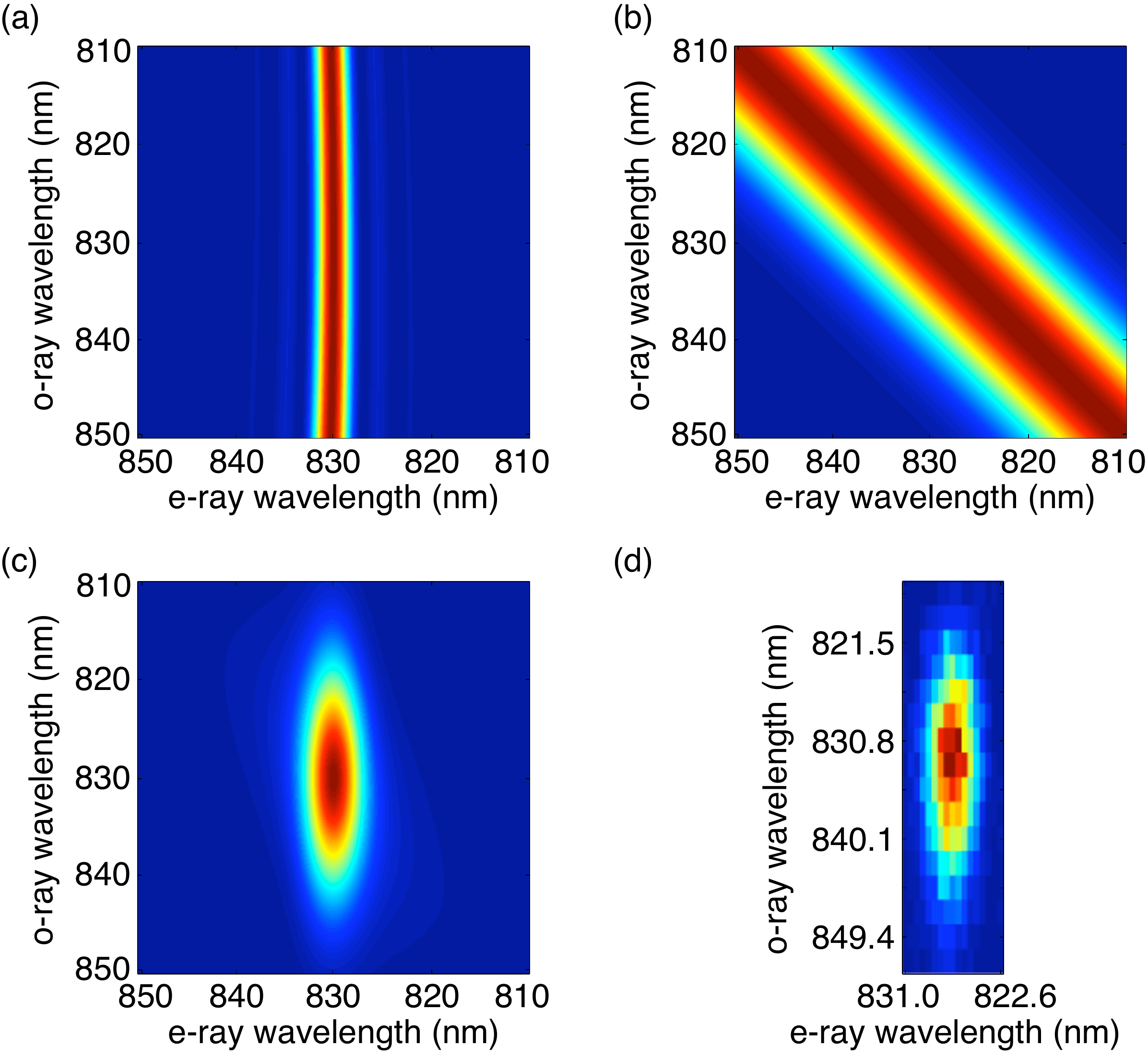}
\caption[Engineering a factorable two-photon state in KDP.]{(Color online) Collinear phasematching intensity, (a), and pump intensity function, (b), for a 5\,mm KDP crystal and an ultrafast pump centered at 415\,nm. The numerically calculated two-photon joint spectral probability distribution, (c), with pump focusing and pair collection as per the experimental parameters. The measured two-photon joint spectral probability distribution is shown in (d). Assuming no variation of the joint spectral phase, the purity of heralded photons with the corresponding amplitude distribution would be 0.98.}
\label{fig:2}
\end{figure}

The two-photon amplitude is given by the product of the pump distribution, $\alpha(\omega_{\text{e}}+ \omega_{\text{o}})$, with the phasematching function of the nonlinear crystal, $\phi(\omega_{\text{e}},\omega_{\text{o}})$:
\begin{equation}
f(\omega_{\text{e}},\omega_{\text{o}}) = \alpha(\omega_{\text{e}}+ \omega_{\text{o}}) \phi(\omega_{\text{e}},\omega_{\text{o}}).
\end{equation}
We select a pump wavelength and crystal for which the pump will propagate with the same group velocity as one of the daughter photons. Arranging the crystal dispersion this way sets $\phi(\omega_{\text{e}},\omega_{\text{o}})$ to be ``vertical'' in frequency space (see  Fig.\,\ref{fig:2}). Using an ultrafast, spectrally broad pump allows this phasematching function to dominate the form of the two-photon amplitude, resulting in a factorable $f(\omega_{\text{e}},\omega_{\text{o}})$. An intuitive explanation for this is that traveling with equal group velocities forces one daughter photon to occupy the same narrow ``time bin'' as the pump pulse and therefore to be in a single broadband mode. Hence the timing jitter of each pair emission event is minimal. To fulfill this group velocity condition it is essential to have the flexibility provided by type-II phasematching where the pump photons decay into a pair in orthogonally polarized modes (e-ray and o-ray).

We have built a pair of sources exhibiting the characteristics required for pure-state generation by utilizing the dispersion characteristics of potassium-dihydrogen-phosphate (KDP). KDP has the property that the group velocity of e-polarized light at 415\,nm is the same as that of o-polarized light at 830\,nm, hence satisfying the group velocity matching condition for factorable states. This makes KDP the ideal crystal for this type of source as firstly it can be pumped by a frequency-doubled femtosecond titanium:sapphire laser, and secondly the daughter photons are easily detectable with room-temperature silicon avalanche photodiodes. As spectral filters are not required, our sources deliver photon pairs at count rates comparable to those in other multi-source downconversion experiments (e.g.,  Refs. \cite{Kaltenbaek2006Experimental-Interference-of-Independent, Lu2007Experimental-entanglement-of-six-photons}) with an order of magnitude less pump power; per crystal, only 40\,mW yields 6000 photon pairs per second.

In addition to being in pure states, we require single photons that are in well-defined spatial modes to implement linear optical gates and distribute across quantum networks. This is achieved by coupling the photon pairs into single-mode optical fiber. To obtain good coupling efficiency, one must properly focus the pump beam into the nonlinear crystals \cite{Ljunggren2005Optimal-focusing-for-maximal} and hence create preferential emission into a single spatial mode. The focusing conditions must be chosen carefully to maintain the factorability of the two-photon state; differences in the phasematching conditions across the range of angles in the pump beam and emission pattern can introduce additional correlations not present in the case of plane-wave pumped KDP (illustrated in Fig.\,\ref{fig:1}). Furthermore, the pump pulses from the second harmonic generation (SHG) system are not transversely homogeneous. Efficient upconversion of the titanium:sapphire laser requires tight focusing of the fundamental beam into the SHG crystal and the resulting range of phasematching angles creates an output beam whose central frequency depends on this angle. The spread of frequencies in this spatially chirped pump beam is mapped to the downconversion crystal so that each phasematching angle sees a different pump wavelength. Careful matching of the sign and magnitude of the spatial chirp of the pump pulses allows the generation of states that have higher factorability than those that would result from homogeneous pump pulses without spatial chirp, while retaining favorable fiber coupling efficiencies. The required focusing and crystal parameters were determined using a numerical model based on equation (\ref{eq:pdc_state}) and including the spatial degree of freedom of the fields in addition to their frequency.

In order to confirm that the chosen source parameters were indeed optimal, the form of the two-photon state from each crystal was verified directly by measuring the joint spectral probability distribution, $|f(\omega_{\text{e}},\omega_{\text{o}})|^{2}$. Photon pairs from one crystal were split and the photons sent to two grating monochromators. The coincidence rates were recorded for pairs of wavelengths to map the joint intensity distribution. Fig.\,\ref{fig:2} shows that the measured probability distribution was indeed factorable and agreed well with the results of the numerical model.

\begin{figure}
\includegraphics[width=0.48 \textwidth]{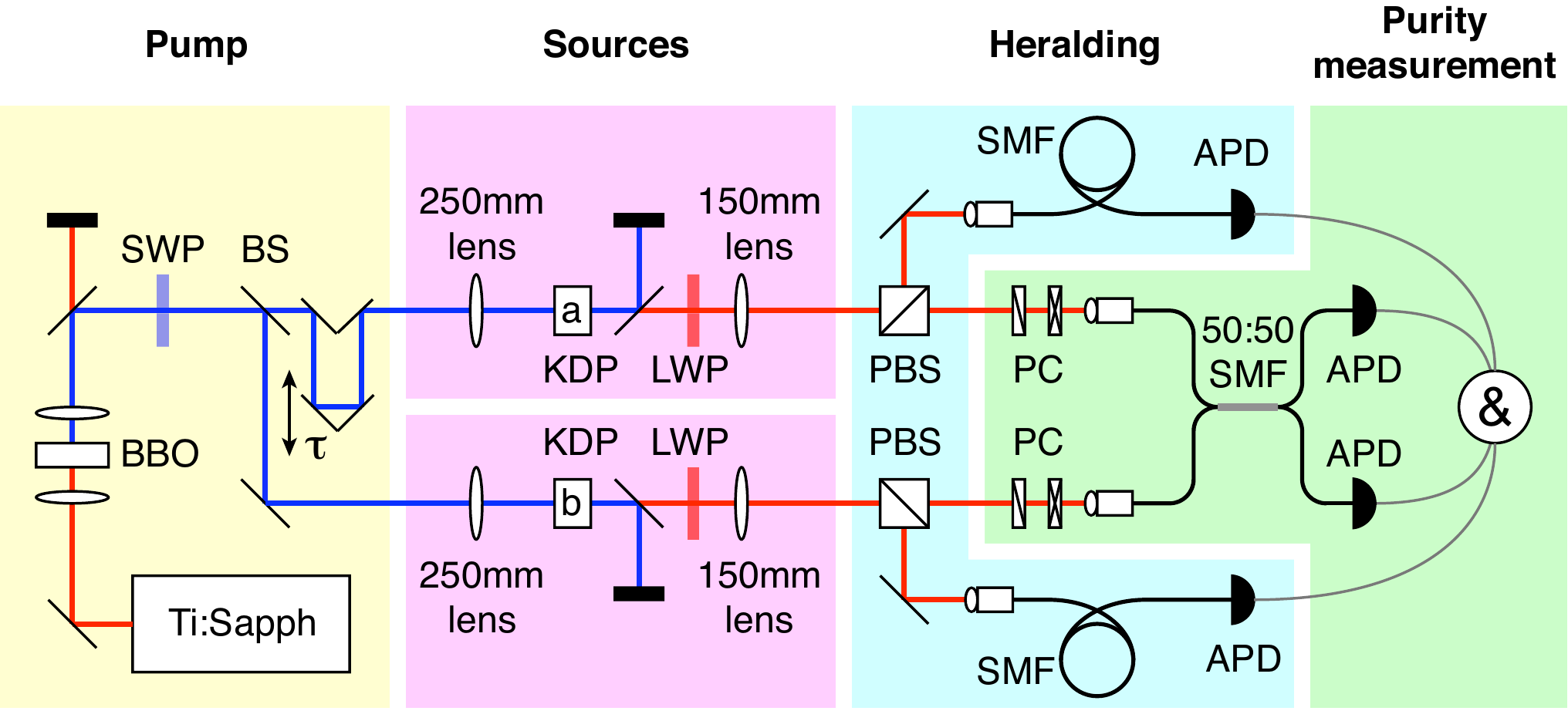}
\caption{(Color online) A titanium:sapphire oscillator (830\,nm, 50\,fs, 76\,MHz) was frequency doubled in a BBO crystal to produce femtosecond pump pulses centered at 415\,nm. After filtering out the fundamental frequency (SWP), these pulses were split at a 50:50 beamsplitter (BS) to give a pair of 40\,mW beams that were focused with two f = 250\,mm SiO$_2$ lenses into two 5\,mm long KDP crystals (labeled a and b) cut for type-II phasematching at 830\,nm. A delay stage ($\tau$) allowed adjustment of the relative arrival time of the two pulses. After the PDC crystals, the pump light was filtered out (LWP) and the photon pairs collimated with f = 150\,mm lenses. The pairs were divided at two polarizing beamsplitters (PBS). The two heralding photons were coupled into single-mode fibers (SMF) with f = 15\,mm aspheric lenses and sent to silicon avalanche photodiodes (APD). The remaining single photons passed through half and quarter waveplates for polarization control (PC) and were directed to a 50:50 single-mode fiber coupler (50:50 SMF) with two APDs on the outputs. Fast electronics then monitored the coincidence rate between the four detectors. No narrowband spectral filters were used.}
\label{fig:3}
\end{figure}

Both the purity and indistinguishability of the heralded single photons were tested concurrently by performing a HOMI between photons from two independent sources \cite{Ou1999Photon-bunching-and-multiphoton}. One photon from each pair was sent directly to a silicon avalanche photodiode (APD) for detection as the herald, and the other directed to a 50:50 single-mode fiber coupler, as shown in Fig.\,\ref{fig:3}. We observed the interference of the two heralded photons by monitoring the rate of fourfold coincidence events at the APDs as a function of the relative optical delay in the two arms. The slow detectors integrated over the whole of each photon wavepacket; therefore the detection was not time-resolved and no temporal filtering could take place.

\begin{figure}
\includegraphics[width=0.3 \textwidth]{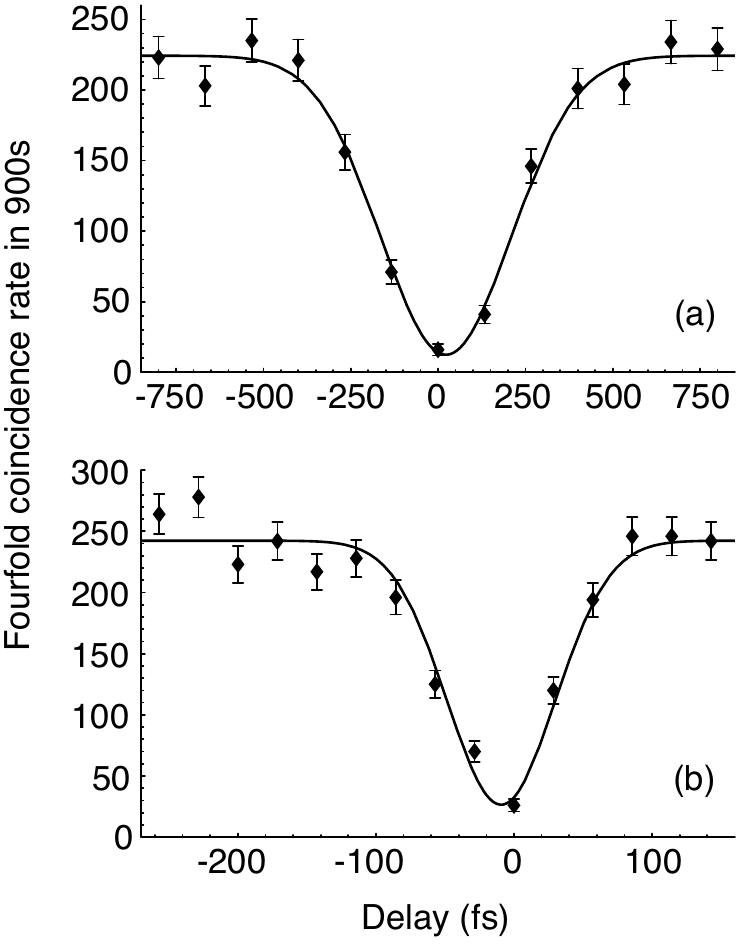}
\caption{Interference of heralded photons from independent sources with no spectral filtering. Interference of  e-ray photons shown in (a) with $V$ = 94.4$\pm$1.6\%, FWHM = 440\,fs, and o-ray photons in (b) with $V$ = 89.9$\pm$3.0\%, FWHM = 92\,fs. The visibility of the o-ray dip is reduced due to the insufficient bandwidth of the 50:50 fiber coupler. Fits are weighted least squares to a Gaussian function and error bars represent Poissonian counting statistics.}
\label{fig:4}
\end{figure}

Heralding with the o-ray photons, we obtained a HOMI dip with 94.4\%$\pm$1.6\% visibility between the e-ray photons (Fig.\,\ref{fig:4}). Upon measuring the two photons' spectra and finding their overlap, the minimum purity was then calculated to be greater than 95\%. The FWHM of the dip was 440\,fs, corresponding to a coherence time of 310\,fs, commensurate with the spectrum of the e-ray photons. In addition, we interfered the broadband o-ray photons, using the e-rays as triggers. This gave a similar visibility (89.9\%$\pm$3.0\%), but the interference dip was much narrower (92\,fs FWHM) showing that the coherence time of the heralded o-ray photons was only 65\,fs.

A benefit of using the e-ray photons to herald the o-ray photons is that a spectral filter in the e-ray arm can be used to increase the heralding efficiency by reducing the background without throwing away any of heralded photons themselves. The filter is chosen with a bandwidth greater than that of the e-ray photons ($\sim$4\,nm) but narrower than the background scatter. We have used this approach to measure detection efficiencies over 25\% --- very high for a pulsed PDC source using bulk optics. Taking into account the quantum efficiency of the signal detector (around 60\%) this gives a heralding efficiency of almost 44\%.

In conclusion, through the use of an ultrafast laser and by group velocity matching one daughter photon to the pump pulse, we have prepared fiber-coupled PDC photon pairs in factorable states. By successfully interfering almost identical photons produced simultaneously in two independent sources, we have shown that from these pairs we obtain single photons with a purity of over 95\% at high heralding efficiencies (up to 44\%). Hence we have demonstrated that lossy spectral filters are no longer required for multiple source pair generation experiments. All the pairs generated are in useable modes, allowing linear optical quantum computing experiments to move beyond post-selecting on the presence of logical-qubit photons. Furthermore, due to its single mode character, our source does not suffer from shot-to-shot fluctuations --- the photons are delivered with precise timing and very low jitter. We therefore expect that the group velocity matching method we have demonstrated will enable a substantial increase in the fidelity of optical quantum gates, currently limited by poor quality sources \cite{Lanyon2007Experimental-demonstration-of-Shors}. The overall problem of non-determinism in the creation of photon pairs could be solved with a broadband quantum memory \cite{Nunn2007Mapping-broadband-single-photon}. This source is of substantial importance for future photonic quantum-enhanced technologies, particularly for techniques requiring high photon number \cite{Ourjoumtsev2006Generating-Optical-Schrodinger} or multiple sources, where spectral filtering is ineffective.

This work was supported by the EPSRC (UK) through the QIP IRC, by the European Commission under the Integrated Project Qubit Applications (QAP) funded by the IST directorate, and by the Royal Society.

\end{document}